\documentclass[english]{article}
\usepackage{times}
\usepackage[T1]{fontenc}
\usepackage[latin1]{inputenc}
\usepackage{geometry}
\usepackage{color}
\usepackage{graphicx}

\makeatletter

\providecommand{\tabularnewline}{\\}

\usepackage{axodraw}
\usepackage{lineno}

\usepackage{babel}
\makeatother
\begin{document}

\title{The single production of the lightest E6 isosinglet quark at the
LHC}

\author{S. Sultansoy\textcolor{black}{$^{1,2}$, G. Unel}$^{3,4}$ and M.
Yilmaz$^{2}$}

\maketitle
\begin{center}\textcolor{black}{$^{1}$Institute of Physics, Academy
of Sciences, Baku, Azerbaijan.}\par\end{center}

\begin{center}\textcolor{black}{$^{2}$Gazi University, Physics Department,
Ankara, Turkey.}\par\end{center}

\begin{center}\textcolor{black}{$^{3}$University of California at
Irvine, Physics Department, USA. }\par\end{center}

\begin{center}\textcolor{black}{$^{4}$CERN, Physics Department, Geneva,
Switzerland.}\par\end{center}

\begin{abstract}
We study the jet associated production of the new quarks predicted
by the $E_{6}$ GUT model at the LHC. Generator level considerations
are made for different mass values of the lightest of the new quarks
to investigate its discovery potential and the prospects for obtaining
its mixing angle to the Standard Model quarks. We find that after
100 fb$^{-1}$ of data taking, it is possible to discover the new
quark with a significance more than 5$\sigma$ up to a mass of 1500
GeV. If no discovery is made, it is possible to constrain the mass
vs quark mixing angle plane.
\end{abstract}

\section{Introduction}

One possible way to unify the electroweak and the strong forces is
embedding the Standard Model (SM) into a larger symmetry group as
the Grand Unified Theories (GUTs) suggest. One possible candidate,
which is also favored as the low energy effective theory of super-strings,
is the $E_{6}$ model \cite{R-e6,R-hewet-rizzo}. The $E_{6}$ model
predicts a number of new particles, among which there is an isosinglet
quark per SM family. The forthcoming LHC accelerator will be the place
to search for the new colored objects in general and in particular
for isosinglet quarks, if their masses are within the LHC kinematical
range. The current experimental limit on the mass of an isosinglet
quark with a $-1/3$ electric charge is $m>199\;$GeV \cite{PDG}.
The distinguishable feature of the isosinglet quarks is the existence
of FCNCs at tree level as opposed to fourth SM family quarks. If a
new quark is observed, this feature can be used to distinguish between
different models. The pair production of the isosinglet quarks was
previously studied for the Tevatron \cite{Rosner} and for the LHC
\cite{details}. The pair production cross section is practically
independent of the mixing between the SM and the isosinglet quarks.
This note addresses the discovery of the isosinglet quarks via their
single production at the LHC and the measurements of the mixing angle
between the new and the SM quarks.

\section{The Model}

We will denote the isosinglet quarks for the first, second, and the
third SM families by letters $D,S,B$, respectively. In accordance
with the SM fermion mass hierarchy, we assume that $m_{D}<m_{S}<m_{B}$
and also that the intra-family mixing dominates the inter-family mixing.
Further details of the model can be found in \cite{details}. The
mixing angle between the $d$ and $D$ quarks will be represented
by the Greek letter $\phi$. Although the current limit on $\phi$
is $|\sin\phi|<0.07$ \cite{details}, in this note we consider a
more conservative value, $\sin\phi=0.045$, for the calculation of
the cross sections and decay widths. For other values, both of these
two quantities can be scaled with a $\sin^{2}\phi$ dependence. The
branching ratios are about 67 \% for $D\rightarrow W\, u$ and about
33 \% for $D\rightarrow Z\, d$. If the masses of the SM quarks are
generated by the Higgs mechanism, the branching ratios will be modified
to incorporate also the $D\rightarrow H\, d$ channel. For $m_{H}\ll m_{D}$,
the branching ratios become 50 \% for CC, 25 \% for NC and 25\% for
the Higgs channel. Further study of the Higgs boson in the context
of $D$ quark decays is in preparation \cite{e6-higgs}.

\section{The Production at the LHC}

\begin{figure}

\caption{Cross section in single D production as a function of D quark mass
for different $\sin\phi$ values.}

\begin{centering}\includegraphics[scale=0.4]{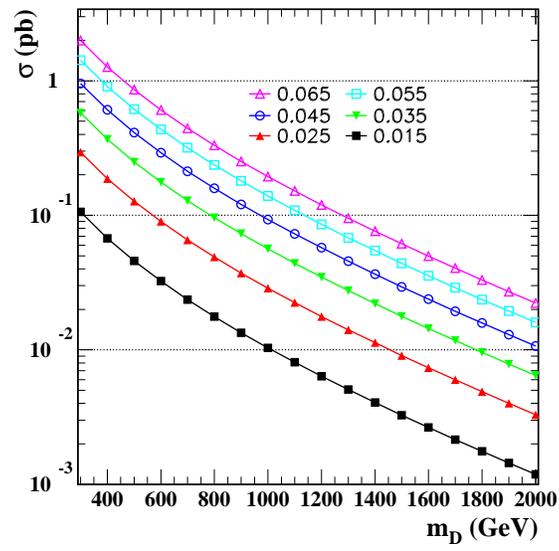}\label{fig:Cross-sections}\par\end{centering}
\end{figure}

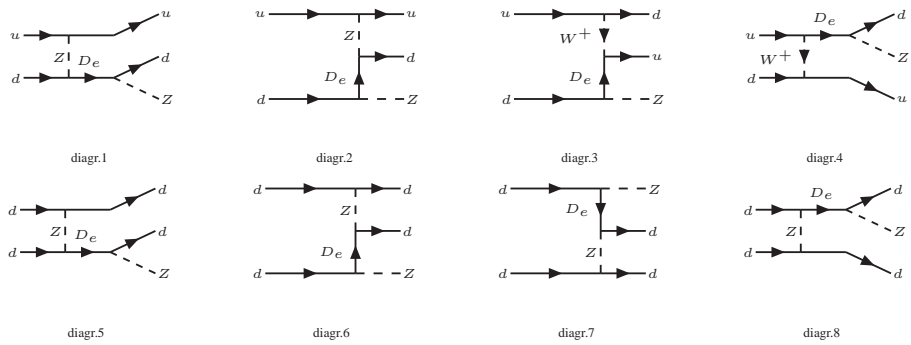
\begin{figure}

\caption{The main tree level signal processes}

\begin{centering}
{
\unitlength=1.0 pt
\SetScale{1.0}
\SetWidth{0.7}      
\tiny    
{} \qquad\allowbreak
\begin{picture}(79,65)(0,0)
\Text(13.0,49.0)[r]{$u$}
\ArrowLine(14.0,49.0)(31.0,49.0) 
\Line(31.0,49.0)(48.0,49.0) 
\Text(66.0,57.0)[l]{$u$}
\ArrowLine(48.0,49.0)(65.0,57.0) 
\Text(30.0,41.0)[r]{$Z$}
\DashLine(31.0,49.0)(31.0,33.0){3.0} 
\Text(13.0,33.0)[r]{$d$}
\ArrowLine(14.0,33.0)(31.0,33.0) 
\Text(39.0,37.0)[b]{$D_{e}$}
\ArrowLine(31.0,33.0)(48.0,33.0) 
\Text(66.0,41.0)[l]{$d$}
\ArrowLine(48.0,33.0)(65.0,41.0) 
\Text(66.0,25.0)[l]{$Z$}
\DashLine(48.0,33.0)(65.0,25.0){3.0} 
\Text(39,0)[b] {diagr.1}
\end{picture} \ 
{} \qquad\allowbreak
\begin{picture}(79,65)(0,0)
\Text(13.0,57.0)[r]{$u$}
\ArrowLine(14.0,57.0)(48.0,57.0) 
\Text(66.0,57.0)[l]{$u$}
\ArrowLine(48.0,57.0)(65.0,57.0) 
\Text(47.0,49.0)[r]{$Z$}
\DashLine(48.0,57.0)(48.0,41.0){3.0} 
\Text(66.0,41.0)[l]{$d$}
\ArrowLine(48.0,41.0)(65.0,41.0) 
\Text(44.0,33.0)[r]{$D_{e}$}
\ArrowLine(48.0,25.0)(48.0,41.0) 
\Text(13.0,25.0)[r]{$d$}
\ArrowLine(14.0,25.0)(48.0,25.0) 
\Text(66.0,25.0)[l]{$Z$}
\DashLine(48.0,25.0)(65.0,25.0){3.0} 
\Text(39,0)[b] {diagr.2}
\end{picture} \ 
{} \qquad\allowbreak
\begin{picture}(79,65)(0,0)
\Text(13.0,57.0)[r]{$u$}
\ArrowLine(14.0,57.0)(48.0,57.0) 
\Text(66.0,57.0)[l]{$d$}
\ArrowLine(48.0,57.0)(65.0,57.0) 
\Text(44.0,49.0)[r]{$W^+$}
\DashArrowLine(48.0,57.0)(48.0,41.0){3.0} 
\Text(66.0,41.0)[l]{$u$}
\ArrowLine(48.0,41.0)(65.0,41.0) 
\Text(44.0,33.0)[r]{$D_{e}$}
\ArrowLine(48.0,25.0)(48.0,41.0) 
\Text(13.0,25.0)[r]{$d$}
\ArrowLine(14.0,25.0)(48.0,25.0) 
\Text(66.0,25.0)[l]{$Z$}
\DashLine(48.0,25.0)(65.0,25.0){3.0} 
\Text(39,0)[b] {diagr.3}
\end{picture} \ 
{} \qquad\allowbreak
\begin{picture}(79,65)(0,0)
\Text(13.0,49.0)[r]{$u$}
\ArrowLine(14.0,49.0)(31.0,49.0) 
\Text(39.0,53.0)[b]{$D_{e}$}
\ArrowLine(31.0,49.0)(48.0,49.0) 
\Text(66.0,57.0)[l]{$d$}
\ArrowLine(48.0,49.0)(65.0,57.0) 
\Text(66.0,41.0)[l]{$Z$}
\DashLine(48.0,49.0)(65.0,41.0){3.0} 
\Text(27.0,41.0)[r]{$W^+$}
\DashArrowLine(31.0,49.0)(31.0,33.0){3.0} 
\Text(13.0,33.0)[r]{$d$}
\ArrowLine(14.0,33.0)(31.0,33.0) 
\Line(31.0,33.0)(48.0,33.0) 
\Text(66.0,25.0)[l]{$u$}
\ArrowLine(48.0,33.0)(65.0,25.0) 
\Text(39,0)[b] {diagr.4}
\end{picture} \ 
}
\par\end{centering}

\begin{centering}
{
\unitlength=1.0 pt
\SetScale{1.0}
\SetWidth{0.7}      
\tiny    
{} \qquad\allowbreak
\begin{picture}(79,65)(0,0)
\Text(13.0,49.0)[r]{$d$}
\ArrowLine(14.0,49.0)(31.0,49.0) 
\Line(31.0,49.0)(48.0,49.0) 
\Text(66.0,57.0)[l]{$d$}
\ArrowLine(48.0,49.0)(65.0,57.0) 
\Text(30.0,41.0)[r]{$Z$}
\DashLine(31.0,49.0)(31.0,33.0){3.0} 
\Text(13.0,33.0)[r]{$d$}
\ArrowLine(14.0,33.0)(31.0,33.0) 
\Text(39.0,37.0)[b]{$D_{e}$}
\ArrowLine(31.0,33.0)(48.0,33.0) 
\Text(66.0,41.0)[l]{$d$}
\ArrowLine(48.0,33.0)(65.0,41.0) 
\Text(66.0,25.0)[l]{$Z$}
\DashLine(48.0,33.0)(65.0,25.0){3.0} 
\Text(39,0)[b] {diagr.5}
\end{picture} \ 
{} \qquad\allowbreak
\begin{picture}(79,65)(0,0)
\Text(13.0,57.0)[r]{$d$}
\ArrowLine(14.0,57.0)(48.0,57.0) 
\Text(66.0,57.0)[l]{$d$}
\ArrowLine(48.0,57.0)(65.0,57.0) 
\Text(47.0,49.0)[r]{$Z$}
\DashLine(48.0,57.0)(48.0,41.0){3.0} 
\Text(66.0,41.0)[l]{$d$}
\ArrowLine(48.0,41.0)(65.0,41.0) 
\Text(44.0,33.0)[r]{$D_{e}$}
\ArrowLine(48.0,25.0)(48.0,41.0) 
\Text(13.0,25.0)[r]{$d$}
\ArrowLine(14.0,25.0)(48.0,25.0) 
\Text(66.0,25.0)[l]{$Z$}
\DashLine(48.0,25.0)(65.0,25.0){3.0} 
\Text(39,0)[b] {diagr.6}
\end{picture} \ 
{} \qquad\allowbreak
\begin{picture}(79,65)(0,0)
\Text(13.0,57.0)[r]{$d$}
\ArrowLine(14.0,57.0)(48.0,57.0) 
\Text(66.0,57.0)[l]{$Z$}
\DashLine(48.0,57.0)(65.0,57.0){3.0} 
\Text(44.0,49.0)[r]{$D_{e}$}
\ArrowLine(48.0,57.0)(48.0,41.0) 
\Text(66.0,41.0)[l]{$d$}
\ArrowLine(48.0,41.0)(65.0,41.0) 
\Text(47.0,33.0)[r]{$Z$}
\DashLine(48.0,41.0)(48.0,25.0){3.0} 
\Text(13.0,25.0)[r]{$d$}
\ArrowLine(14.0,25.0)(48.0,25.0) 
\Text(66.0,25.0)[l]{$d$}
\ArrowLine(48.0,25.0)(65.0,25.0) 
\Text(39,0)[b] {diagr.7}
\end{picture} \ 
{} \qquad\allowbreak
\begin{picture}(79,65)(0,0)
\Text(13.0,49.0)[r]{$d$}
\ArrowLine(14.0,49.0)(31.0,49.0) 
\Text(39.0,53.0)[b]{$D_{e}$}
\ArrowLine(31.0,49.0)(48.0,49.0) 
\Text(66.0,57.0)[l]{$d$}
\ArrowLine(48.0,49.0)(65.0,57.0) 
\Text(66.0,41.0)[l]{$Z$}
\DashLine(48.0,49.0)(65.0,41.0){3.0} 
\Text(30.0,41.0)[r]{$Z$}
\DashLine(31.0,49.0)(31.0,33.0){3.0} 
\Text(13.0,33.0)[r]{$d$}
\ArrowLine(14.0,33.0)(31.0,33.0) 
\Line(31.0,33.0)(48.0,33.0) 
\Text(66.0,25.0)[l]{$d$}
\ArrowLine(48.0,33.0)(65.0,25.0) 
\Text(39,0)[b] {diagr.8}
\end{picture} \ 
}\label{fig:The-main-diags}\par\end{centering}
\end{figure}

We implemented the Lagrangian for the signal into a tree level Monte
Carlo generator, CompHep v4.4.3 \cite{R-calchep}. For both the signal
and the background studies, the contributions from sea quarks were
also considered. The parton distribution function utilized was CTEQ6L1
and the QCD scale was set to be the mass of the $D$ quark for both
signal and background processes. The cross section for single production
of the $D$ quark for its mass up to 2 TeV and for various mixing
angles is given in figure \ref{fig:Cross-sections}. The main tree
level signal processes with their neutral current decays are given
in figure \ref{fig:The-main-diags}. The remaining processes originating
from the sea quarks contribute about 20 percent to the total signal
cross section.

\begin{figure}

\caption{The transverse momentum distribution of the parton with the highest
energy for the signal case.}

\begin{centering}\includegraphics[scale=0.5]{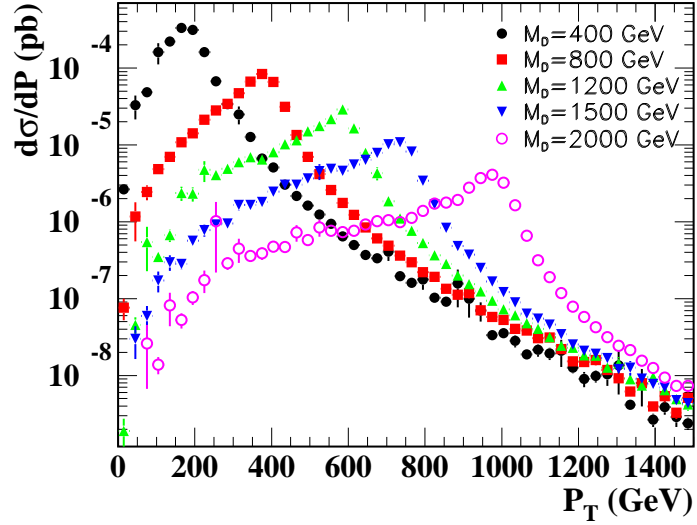}\label{fig:si-pt}\par\end{centering}
\end{figure}

\begin{figure}

\caption{The transverse momentum distribution of the parton with the highest
energy for the background case.}

\begin{centering}\includegraphics[scale=0.5]{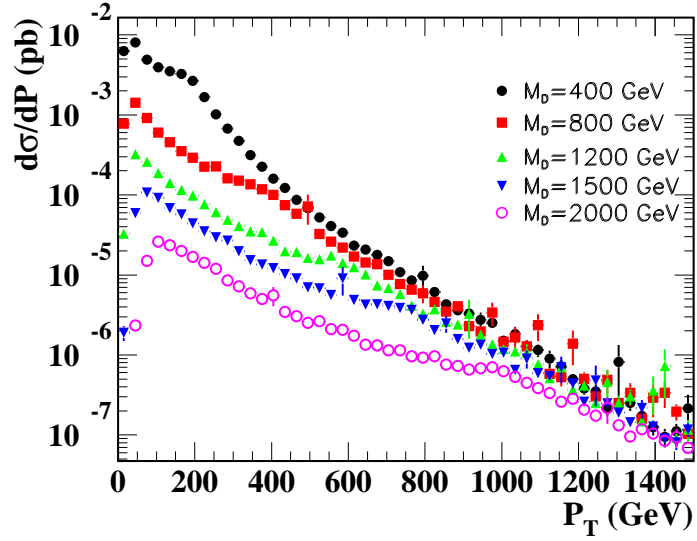}\label{fig:bg-pt}\par\end{centering}
\end{figure}

Although this note considered a generator level study, various parameters
of the ATLAS detector \cite{R-atlas-tdr} such as the barrel calorimeter
geometrical acceptance, minimum angular distance for jet separation
and minimum transverse momentum for jets\cite{tdaq-tdr} were taken
into account. Four mass values (400, 800,1200, 1500 and 2000 GeV)
were studied to investigate the mass dependence of the discovery potential
for this channel. The cuts common to all considered mass values are:

\begin{eqnarray*}
P_{Tp} & > & 15\,\hbox{GeV}\\
|\eta_{p}| & < & 3.2\\
|\eta_{Z}| & < & 3.2\\
R_{p} & > & 0.4\\
M_{Zp} & = & M_{D}\pm20\,\hbox{GeV}\end{eqnarray*}

where $p$ stands for any parton; $R$ is the cone separation angle
between two partons; $\eta_{p}$ and $\eta_{Z}$ are pseudorapidities
of a parton and $Z$ boson respectively; and $P_{Tp}$ is the parton
transverse momentum. The main property that allows discriminating
between the signal and the background is the transverse momentum distribution
of the parton with the highest energy. In figure \ref{fig:si-pt},
the transverse momentum distributions of the partons that survive
the common cuts are shown for the signal case. One can observe that
the distribution peaks at about one half of the $D$ quark mass value.
The same distribution for the background is given in figure \ref{fig:bg-pt},
which shows rapidly decreasing differential cross section as a function
of parton $P_{T}$ since the final state partons are originating directly
from the initial partons.

\subsection{The discovery potential}

\begin{center}%
\begin{figure}

\caption{Dependence of the signal significance on hard jet $P_{T}$ for different
$D$ quark mass values}

\begin{centering}\includegraphics[scale=0.7]{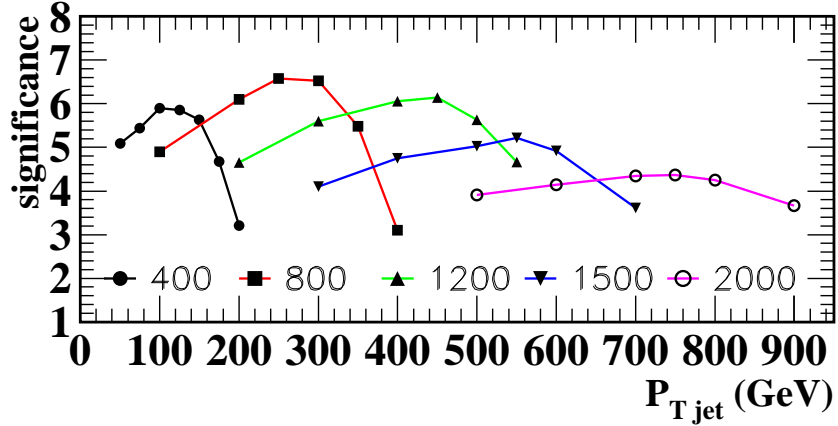}\label{fig:signif-scan}\par\end{centering}
\end{figure}
\par\end{center}

The results from the scanning of the signal significance for the hard
jet $P_{T}$ are given in figure \ref{fig:signif-scan} for $D$ quark
mass values under study. For each case, the optimal value is found
by maximizing the significance ($S/\sqrt{B}$) and it is used for
calculating the effective cross sections presented in table \ref{tab:effective-sigmas}.
To obtain the actual number of events for each mass value, the $e\bar{e}$
and $\mu\bar{\mu}$ decays of the $Z$ boson were considered for simplicity
of reconstruction. Table \ref{tab:Expected-events} contains the expected
number of reconstructed events for both signal and background for
100 fb$^{-1}$of data taking. Although the lepton identification and
reconstruction efficiencies are not considered, one can note that
the statistical significance at $m_{D}=$1500 GeV, is above 5$\sigma$
after one year of nominal luminosity run.

\begin{table}

\caption{The signal and background effective cross sections before the $Z$
decay and after the optimal cuts, together with the $D$ quark width
in GeV for each considered mass. }

\begin{centering}\begin{tabular}{|c|c|c|c|c|c|}
\hline 
$M_{D}$(GeV)&
400&
800&
1200&
1500&
2000\tabularnewline
\hline
\hline 
$\Gamma$(GeV)&
0.064&
0.51&
1.73&
3.40&
8.03\tabularnewline
\hline 
Signal (fb)&
100.3&
29.86&
10.08&
5.09&
1.92\tabularnewline
\hline 
Background (fb)&
2020&
144&
18.88&
6.68&
1.36\tabularnewline
\hline
optimal $P_{T}$ cut&
100&
250&
450&
550&
750\tabularnewline
\hline
\end{tabular}\label{tab:effective-sigmas}\par\end{centering}
\end{table}

\begin{table}

\caption{Expected number of reconstructed events using electron and muon decays
of the $Z$ boson and signal significance for 100 fb$^{-1}$ of data
taking.}

\begin{centering}\begin{tabular}{|c|c|c|c|c|c|}
\hline 
$M_{D}$(GeV)&
400&
800&
1200&
1500&
2000\tabularnewline
\hline
\hline 
Signal Events&
702&
209&
71&
36&
13.5\tabularnewline
\hline 
Background Events&
14000&
1008&
132&
47&
9.5\tabularnewline
\hline 
Signal significance&
5.9&
6.6&
6.1&
5.2&
4.37\tabularnewline
\hline
\end{tabular}\label{tab:Expected-events}\par\end{centering}
\end{table}

\subsection{Extracting the quark mixing angle }

\begin{figure}

\caption{3$\sigma$ exclusion curves for 10, 100, 300, 1000 fb $^{-1}$ integrated
luminosities are shown from top to down. \label{cap:Determination-of-sinphi}}

\begin{centering}\includegraphics[scale=0.45]{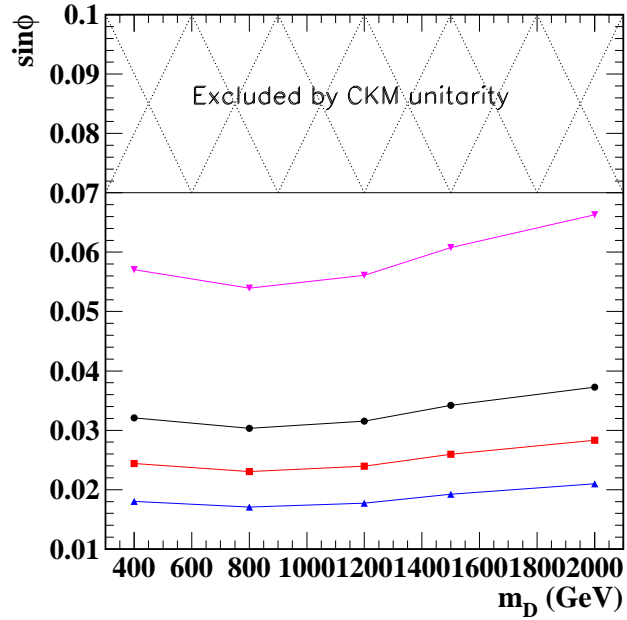}\par\end{centering}
\end{figure}

As mentioned before, the pair production cross section is almost independent
of the $D-d$ quarks mixing angle, therefore we will use the single
production discovery results given in the previous section to investigate
this angle. In the event of a discovery in the single production case,
the mixing angle can be obtained directly. If no discoveries are made,
then the limit on the cross section can be converted to a limit curve
in the $D$ quark mass vs mixing angle plane. Therefore we calculated
the angular reach for a 3$\sigma$ signal by extrapolating to other
$\sin\phi$ values. Figure~\ref{cap:Determination-of-sinphi} gives
the mixing angle versus $D$-quark mass plane and the 3$\sigma$ reach
curves for different integrated luminosities ranging from 10 fb$^{-1}$
to 1000 fb$^{-1}$, which correspond to one year of low luminosity
LHC operation and one year of high luminosity super-LHC operation
respectively. The hashed region in the same plot is excluded using
the current values of the CKM matrix elements. One should note that,
this channel allows reducing the current limit on $\sin\phi$ by half
in about 100 fb$^{-1}$ run time.

\section{The Results and Conclusions}

The process of single production of the $E_{6}$ isosinglet quarks
could essentially enhance the discovery potential if $\sin\phi$ exceeds
0.02. For example, with 300 fb$^{-1}$ integrated luminosity, the
3$\sigma$ discovery limit is $m_{D}=2000\,$GeV, if $\sin\phi=0.03$.
It should also be noted that for pair production the 3$\sigma$ discovery
limit was found to be about 900 GeV, independent of $\sin\phi$. If
ATLAS discovers an 800 GeV $D$ quark via pair production, single
production will give the opportunity to confirm the discovery and
measure the mixing angle if $\sin\phi>0.03.$ The FCNC decay channel
analyzed in this paper is specific for isosinglet down type quarks
and gives the opportunity to distinguish it from other models also
involving additional down type quarks, for example the fourth SM family.
Therefore, the possible discovery of isosinglet quarks at the LHC
would validate the $E_{6}$ as a GUT group.

\subsection*{Acknowledgments}

S.S. and M.Y. acknowledge the support from the Turkish Atomic Energy
Authority and the Turkish State Planning Organization under the contract
DPT2006K-120470. G.U.'s work is supported in part by U.S. Department
of Energy Grant DE FG0291ER40679. The authors would like to thank
B. Golden, M. Karagoz and A. Lankford for useful comments.

\end{document}